\documentstyle[aps,preprint,prb]{revtex}
\tighten
\begin{document}
\draft
\title{Magnetotransport in a pseudomorphic
GaAs/Ga$_{0.8}$In$_{0.2}$As/Ga$_{0.75}$Al$_{0.25}$As
heterostructure with a Si $\delta$-doping layer}

\author{M. van der Burgt\cite{ookinleuven}}
\address{University of Oxford, Department of Physics,
Clarendon Laboratory,\\ Parks Road, Oxford OX1 3PU}

\author{V. C. Karavolas and  F. M. Peeters}
\address{University of Antwerp (UIA), Department of
Physics,
B-2610 Antwerpen}

\author{J. Singleton and R. J. Nicholas}
\address{University of Oxford, Department of Physics,
Clarendon Laboratory,\\ Parks Road, Oxford OX1 3PU}

\author{F. Herlach}
\address{Katholieke Universiteit Leuven, Department of
Physics,\\ Celestijnenlaan 200D, B-3001 Leuven}

\author{J. J. Harris}
\address{Interdisciplinary Research Centre for
Semiconductor
Materials, Imperial College,\\ Prince Consort Road,
London SW7
2BZ}

\author{M. Van Hove and G. Borghs}
\address{IMEC vzw, Kapeldreef 75, B-3001 Leuven}

\date{\today}

\maketitle

\begin{abstract}

Magnetotransport properties of a pseudomorphic GaAs/\-
Ga$_{0.8}$In$_{0.2}$As/\-Ga$_{0.75}$Al$_{0.25}$As
heterostructure
are investigated in pulsed magnetic fields up to 50
T and at temperatures of $T=$1.4~K and 4.2~K.
The structure studied consists of a Si $\delta$-layer
parallel
to a Ga$_{0.8}$In$_{0.2}$As quantum well (QW). The dark
electron density of
the structure is $n_e=1.67\times 10^{16}\ \rm m^{-2}$. By
illumination the density  can be increased  up to a
factor of
4;  this way the second subband in the
Ga$_{0.8}$In$_{0.2}$As QW can become
populated as well as  the Si $\delta$-layer. The presence
of
electrons in  the
$\delta$-layer results in drastic changes in the
transport
data, especially at magnetic fields beyond 30\ T.  The
phenomena observed are interpreted as: 1) magnetic
freeze-out
of carriers in the $\delta$-layer when a low density of
electrons is present in the
$\delta$-layer, and 2) quantization of the electron
motion in
the two dimensional electron gases in both the
Ga$_{0.8}$In$_{0.2}$As QW and
the Si $\delta$-layer in the case of high densities.
These
conclusions are corroborated by the numerical results of
our
theoretical model.
We obtain a satisfactory agreement between model and
experiment.
\end{abstract}

\narrowtext

\section{INTRODUCTION}

Since the discovery of the two dimensional electron gas
(2DEG)
at the interface of a
GaAs/Ga$_{1-x}$Al$_{x}$As heterojunction,\cite{stormer79}
the
physical properties of 2DEGs in III-V semiconductors have
been
intensively investigated.
One of the basic requirements for fast electronic devices
based on such semiconductor 2DEGs is a high conductivity.
In modulation doped GaAs/Ga$_{1-x}$Al$_{x}$As
heterostructures
\cite{dingle} grown by molecular beam epitaxy (MBE)
electron
mobilities as high as $1000\ \rm m^2/Vs$ can be achieved
at temperatures below 4 K.\cite{foxon89}
However, the two dimensional (2D) electron densities in
these
high mobility systems are usually
not much larger than a few times  $10^{15}\ \rm m^{-
2}$.\cite{foxon89,pfeiffer89}  At high temperatures the
mobility in 2D GaAs based systems is strongly limited by
optical phonon
scattering.\cite{lee83,walukwiecz84,ohno88,luo}
Therefore, high electron densities are needed in order to
achieve high room temperature conductivities.  Much
greater densities of 2D electrons then in GaAs/Ga$_{1-
x}$Al$_{x}$As structures can
be accommodated  in GaAs/\-Ga$_{1-x}$In$_{x}$As/\-Ga$_{1-
y}$Al$_{y}$As systems.\cite{rosenberg} This is due to the
large  conduction band
offset,\cite{adachi82,adachi85,tiwari92} which results in
larger  confinement energies for the 2D electrons.  When
the Ga$_{1-x}$In$_{x}$As layer is thin enough the
lattice mismatch between Ga$_{1-x}$In$_{x}$As and Ga$_{1-
y}$Al$_y$As is
accommodated by the Ga$_{1-x}$In$_{x}$As layer without
introducing misfit
dislocations,\cite{andersson87,fritz87} resulting in a
pseudomorphic structure.

 The electron concentrations in GaAs/\-Ga$_{1-
x}$In$_{x}$As/\-Ga$_{1-y}$Al$_{y}$As structures can be
further increased by making use of a $\delta$-function
like
doping profile \cite{zrenner85,schubert87} instead of a
uniformly Si doped
Ga$_{1-y}$Al$_{y}$As layer. This also results in an
enhanced
mobility \cite{cunningham88,schubertbook} as compared to
modulation doped structures with uniformly doped layers.
In this way, electron densities higher than $1.5\times
10^{16}\ \rm
m^{-2}$ and low temperature mobilities of $\mu \approx3$-
-$4 \ \rm m^2/Vs$ can be achieved. The high densities and
mobilities, and the resulting
high room temperature conductivities in these systems
makes
them attractive for use in high performance high-electron
mobility transistors (HEMTs).\cite{mimura80}

The presence of DX centers in Si doped Ga$_{1-
x}$Al$_{x}$As
with $x \geq 0.20$ is responsible for the persistent
photoconductivity in this
material.\cite{theis,mooney,schubertbook2} The ionisation
of
the DX centers after illumination can lead
to a parallel conducting channel in modulation doped
heterostructures as carriers do not recombine with the
ionized
DX centers.\cite{luryi84} Since the carriers in the
parallel
layer have a low mobility compared to the 2D electrons in
the
quantum well (QW) or at the heterojunction interface,
parallel
conduction is usually regarded as an undesirable effect
which degrades the performance of a
HEMT.\cite{tian91,liu91}
 Most studies of parallel conduction are therefore mainly
concerned with the question of how to avoid
it,\cite{kudo94}
or how to extract mobilities and densities of the
different
conducting channels.\cite{syphers86,jharris91}
 In this paper we will show that conduction in a
$\delta$-layer parallel with a Ga$_{0.8}$In$_{0.2}$As QW
can give rise to
interesting effects in high magnetic fields, which are
due to
the
interplay between the 2D carriers in the $\delta$-layer
and
the Ga$_{0.8}$In$_{0.2}$As QW, and the 2D character of
the carriers in the Si
$\delta$-layer in the Ga$_{1-x}$Al$_{x}$As.
The 2D character of electrons in a Si $\delta$-layer in
GaAs
was first demonstrated by Zrenner {\it et
al.}\cite{zrenner85}
In
high magnetic fields a metal-insulator transition due to
magnetic freeze-out was observed in $\delta$-doped GaAs
systems.\cite{zrenner88,ye89}

We report transport measurements in pulsed
magnetic fields up to 50 T on a GaAs/\-
Ga$_{0.8}$In$_{0.2}$As/\-Ga$_{0.75}$Al$_{0.25}$As
structure with a Si
doped $\delta$-layer in the Ga$_{0.75}$Al$_{0.25}$As
layer. The initial data, together with a brief
qualitative description were presented in two previous
reports.\cite{chiba,boston}  In this paper we want to
give a description of the data together with a full
theoretical modelling over the whole experimental field
range.

In the system studied, the density of conduction
electrons can be increased by
illuminating the sample with red light.
 At the lowest electron densities  the quantized Hall
effect
(QHE) is observed, with one subband
occupied in the Ga$_{0.8}$In$_{0.2}$As QW. As the density
is increased the
second subband becomes populated.
 Further increases of the electron density result in
parallel
conduction in the $\delta$-layer which leads to drastic
changes in the transport coefficients $\rho_{xx}$ and
$\rho_{xy}$,
especially at fields
above 30 T. These are attributed to magnetic freeze-out
in the
$\delta$-layer.  With the $\delta$-layer parallel to a
conducting 2DEG in the
Ga$_{0.8}$In$_{0.2}$As QW, it is possible to study
magnetic freeze-out
without the problems of a diverging resistance in the
$\delta$-layer. At the highest electron concentrations
the QHE
is
observed both in the Ga$_{0.8}$In$_{0.2}$As QW and the Si
$\delta$-layer.

The experimental arrangements are described in Sec.
\ref{expsetup}. In Sec. \ref{subbands} a brief discussion
of the transport data is presented using solutions of the
coupled
Poisson and Schr\"odinger equations, which gives a
qualitative
picture of the different subbands in the
Ga$_{0.8}$In$_{0.2}$As QW and in the
Si $\delta$-layer.
 The results of the two-carrier analysis of the low field
transport data (Sec. \ref{lowmf}) are used to model the
conductivity tensor for high magnetic fields (Sec.
\ref{highmf}). The theoretical description of the high
field data combines some existing theoretical models,
resulting in an accurate description of the measured
$\rho_{xx}$ and $\rho_{xy}$ data for the whole field
range up to 50 T.

\section{EXPERIMENTAL DETAILS}\label{expsetup}

The transport experiments described in this paper were
performed on
a modulation doped pseudomorphic GaAs/\-
Ga$_{0.8}$In$_{0.2}$As/\-Ga$_{0.75}$Al$_{0.25}$As
heterostructure grown
by
molecular beam epitaxy. The structure of the sample is
shown
in Fig. \ref{sample}.
It consists of a  13 nm thick Ga$_{0.8}$In$_{0.2}$As
layer \cite{kapitan} grown on a GaAs
substrate with a GaAs buffer, followed by a 5 nm wide
Ga$_{0.75}$Al$_{0.25}$As spacer
layer, a Si $\delta$-layer with a Si concentration of
$5\times
10^{16}\ \rm m^{-2}$, a 30 nm thick Si doped ($5\times
10^{23}
\ \rm m^{-3}$) Ga$_{0.75}$Al$_{0.25}$As layer and  $n^+$-
GaAs
cap layer of approximately  5 nm.
The samples were mesa etched into 55 $\mu$m wide Hall
bars.
Two samples from two wafers with a nominally  identical
structure were studied (only the thickness of the $n^+$-
GaAs
cap layer was
slightly different); the results from the two samples
were
almost identical and so  only the results of one will be
discussed in this paper.

Through the persistent photoconductivity effect in the
Ga$_{0.75}$Al$_{0.25}$As layer, the electron density in
the
sample studied can be increased by illumination with a
red
LED.  The electron density is $n_e=1.67\times 10^{16}\
\rm m^{-2}$ and
the
mobility is $\mu=3.2\ \rm m^2/Vs$, when the sample is
cooled
to
4.2 K in the dark.
After sufficient illumination the total electron density
can
be increased by a factor of almost 4.  The structure of
the
conduction band of the sample under different
illumination
conditions is shown in Fig. \ref{ec} and is further
discussed
in the next section.

The experiments were done in a liquid helium bath
cryostat at
temperatures of $T$=1.4 and  4.2 K in the  pulsed
magnetic
field facilities at Oxford \cite{oxford} and
Leuven.\cite{leuven} In both installations the magnetic
field
is
generated by discharging a capacitor bank into a
reinforced
solenoid immersed in liquid nitrogen; maximum fields
between
45
and 50 T can be achieved in both systems.
  The pulse shape is half a period of a damped sine wave
which
reaches its peak after 4--8 ms; the total pulse duration
can
be varied in the range 15--40 ms.

Both the Hall effect $\rho_{xy}$ and the longitudinal
magnetoresistance  $\rho_{xx}$ were measured using dc
currents in
the range 1--10 $\mu \rm A$.  Due to its high carrier
density,
the high field magnetoresistance of the sample is
relatively
low ($\approx 30\ \rm k\Omega$ in fields up to 50 T)
compared
to high mobility GaAs/Ga$_{1-x}$Al$_{x}$As
heterojunctions
which have mostly a low carrier density. Therefore the
distortions seen in transport measurements on a 2DEG in
transient magnetic fields described in Ref.\
\onlinecite{wiggelpaper} are negligible for the
data taken during the down sweep of the pulse.
The down sweep is generally a few milliseconds longer
than the
up sweep, so that the signal to noise ratio of the
data taken during the down sweep is usually slightly
better.
Altough there is usually a good agreement between the
data
from the up and the down sweep, for clarity only data
taken
during the down sweep are shown in the following
sections.
 The voltages measured on the sample are amplified by a
PARC/EG\&G differential amplifier Model 113 or 5113  and
recorded by a fast digital transient recorder.  The data
are
then transferred to a personal computer. The magnetic
field is
determined by measuring
the induced voltage in a pick-up coil with an accurately
known
area mounted in the vicinity of the sample.
More details about transport measurements in pulsed
magnetic
fields can be found in Ref.\ \onlinecite{mallett}.

\section{THE BAND STRUCTURE}\label{subbands}

In order to get a qualitative picture of how the
different
subbands in the Ga$_{0.8}$In$_{0.2}$As and in the
$\delta$-layer behave when
the carrier density is increased by illumination, we
calculated the subband structure by solving the coupled
Poisson-Schr\"odinger equations  self-
consistently.\cite{afs,sterndassarma}
The persistent photo-conductivity effect can be
mimicked in these calculations by reducing the ionisation
energy of the Si donors. A smaller ionisation energy
results
in a higher density of ionized donors and consequently
a higher electron density in the conduction band.
Since the shape and the population of the conduction band
influences the magnetic field dependence of the transport
coefficients $\rho_{xx}$ and $\rho_{xy}$, this section
also summarizes our
basic conclusions about the transport data.

An overview of the changes in the conduction band with
illumination is shown in Fig. \ref{ec}. Figure
\ref{ec}(a)
shows the band structure in the unilluminated sample:
only the
lowest subband $E_1$ is occupied. Under these
circumstances we
observe the quantized Hall effect due to the two
dimensional
electron gas (2DEG) in this subband.

A small dose of illumination results in the occupation of
the
second subband $E_2$ of the Ga$_{0.8}$In$_{0.2}$As
quantum well
(Fig. \ref{ec}(b)). As we will see in the next section
this
clearly shows up
in
 the low field $\rho_{xx}$ data as a beat of two
frequencies in the
Shubnikov-de Haas (SdH) oscillations.
 As the Fermi level $E_F$ still lies below the lowest
states
in the
$\delta$-layer, no carriers are found in this layer.

A further increase in carrier density results in the
occupation of the  Si impurity layer when the lowest
state in
the $\delta$-layer $E_\delta$ falls below the Fermi
energy
(Fig. \ref{ec}(c)); at this point no obvious sign of the
second subband
$E_2$ in the Ga$_{0.8}$In$_{0.2}$As remains in the
transport data. The reason
for this is that $E_\delta$ falls below $E_2$ and so this
subband is consequently
depopulated. As will be shown in Sec. \ref{highmf}, at
low
densities in the $\delta$-layer ($<1\times 10^{16}\ \rm
m^{-2}$) and sufficiently high magnetic fields ($>20\ \rm
T$),
a metal-insulator transition occurs in the $\delta$-
layer.

The bandstructure at saturated or nearly saturated
densities
is illustrated in Fig. \ref{ec}(d). Two subbands are
again
occupied in the Ga$_{0.8}$In$_{0.2}$As quantum well. Due
to the small number
of carriers in the second subband  as compared to these
in the
first
subband  and in the $\delta$-layer,
the contribution of the second subband to the
conductivity is
small and
in fact does
not show up in the transport data.
 The electron density in the $\delta$-layer is then
sufficiently large such that we can treat these electrons
as a
second 2DEG. The transport data then show a superposition
of the QHE from both the Ga$_{0.8}$In$_{0.2}$As quantum
well and the Si $\delta$-
layer.

\section{SMALL MAGNETIC FIELDS}\label{lowmf}

\subsection{Theory}\label{sdhosc}

In the low magnetic field range the SdH oscillations in
the
resistivity $\rho_{xx}$ of a single subband can be
described by the
model of Isihara and Smr\u{c}ka,\cite{isihara} which was
improved by Coleridge {\it et al.}\cite{coleridge} by the
introduction of two distinct relaxation times.
Since at low fields the cyclotron orbit of an electron
has a
large spatial extent this model takes into account
multiple
scattering.
The
calculation of the conductivity is based on a constant
density
of states (DOS)  $g_0= m^*/\pi\hbar^2$ ($m^*$ is the
effective mass and $\hbar$ is Planck's constant over
$2\pi$) with a sinusoidal
oscillating part superimposed.
The oscillating part of the DOS reflects the onset of
Landau
quantization and leads to the SdH oscillations in the
magnetoconductivity.

The conductivity tensor given by Coleridge {\it et
al.}\cite{coleridge} can be inverted to give the
resistivities $\rho_{xx}$ and $\rho_{xy}$. For small
magnetic fields the resistivity  $\rho_{xx}$ can be
written as\cite{afs,coleridge}
\begin{equation}\label{rhosdh}
\rho_{xx}= \rho_0\left[1-4e^{-\pi /\omega_c\tau_q}
\frac{X}{\sinh X}\cos(\frac{2\pi
E_F}{\hbar\omega_c})\right].
\end{equation}
In this expression  $\rho_0$ is the zero-field
resistivity, $\omega_c = eB/m^{*}$ is the cyclotron
frequency, $E_F=\pi\hbar^2 n_e/m^{*}$ is the Fermi
energy, $X=2\pi^2 k_B
T/\hbar\omega_c$ and $k_B$ is
Boltzmann's constant.
Both the {\em transport} scattering time $\tau_t= m^*
/\rho_0 e^2 n_e$ and
the {\em single particle} scattering time $\tau_q$ are
present in this expression.
In modulation doped 2D systems
these  can differ by more than an order of
magnitude.\cite{paalanen,harrang,bockelmann,kusters}
The zero-field resistivity $\rho_0$ is determined by the
classical or transport scattering time $\tau_t$, while
the single particle relaxation time or quantum mechanical
scattering time $\tau_q$ is present in
the oscillatory part of the DOS. The transport scattering
rate, $1/\tau_t$, contains no contribution from forward
scattering and small angle scattering receives a very
small weight, as these scattering events have a very
limited effect on the electron drift velocity.  In the
single particle scattering rate $1/\tau_q$, however,
every scattering event is equally important.

The single particle scattering time $\tau_q$ can be
determined from the decay of the amplitude of the SdH
oscillations with decreasing magnetic
field.
The temperature dependence of the oscillations is given
by the
factor $X/\sinh X$ in Eq.~(\ref{rhosdh}).
Since $\omega_c$ contains the effective mass $m^*$,
measuring
the amplitude of the oscillations as a function of
temperature
allows the determination of  $m^*$.  The cosine factor in
Eq. (\ref{rhosdh}) shows that the SdH oscillations are
periodic in
inverse magnetic field.
The carrier density of the 2D subband is determined by
measuring the period.

This model is valid for low and intermediate fields such
that
$\omega_c \tau_q \leq 1$.
For larger magnetic fields localization of the electrons
away from the center of the Landau level
\cite{aoki,joynt} starts to play an
important role,  so that the above model is no longer
valid.\cite{coleridge,englert2}

\subsection{Experimental results}

Densities, transport mobilities and transport scattering
times
can be obtained from the low field $B\rightarrow 0$
limit of the transport coefficients $\rho_{xx}$ and
$\rho_{xy}$. We measured the Hall
density $n_H$ from the slope of the Hall resistance at
low
magnetic fields
\begin{equation}\label{eqnhall}
\rho_{xy} (B\rightarrow 0)=B/n_H e.
\end{equation}
The Hall mobility $\mu_H$ is then defined by
\begin{equation}
\rho_{xx} (B=0)=\rho_0 = 1/n_H \mu_H e.
\end{equation}
For a single carrier system the Hall density is equal to
the
total density and the Hall mobility is given by $\mu_H =
e\tau_t / m^\ast$. When more then one type of carrier is
present $\mu_H$ represents a weighed average mobility
(see Sec. \ref{sectwoband}).

The 4.2 K resistivities $\rho_{xx}$ and $\rho_{xy}$ in
fields up to 5 T
are shown in
Fig. \ref{5tshots} for four different densities.
After cooling in the dark, $n_H=1.67\times 10^{16}
\rm~m^{-2}$ and $\rho_{xx}$ shows a single series of SdH
oscillations on a constant background (Fig.
\ref{5tshots}(a)). The Hall effect $\rho_{xy}$ is a
straight line on which plateaux start to  develop
above 2.5 T.  This behavior of $\rho_{xx}$ and
$\rho_{xy}$  indicates
that there is only one type
of carrier involved in the conduction process at this
density.\cite{coleridge,bockelmann,stormer}

At a Hall density of $n_H=1.75\times 10^{16}\ \rm m^{-
2}$, a
beat in the SdH oscillations is seen in $\rho_{xx}$ (Fig.
\ref{5tshots}(b)), indicating the population of the
second subband. The occupation of the second subband
results in a new scattering channel, giving rise to
intersubband scattering.\cite{stormer} This causes an
intermodulation of
the two subbands producing the
beat.\cite{coleridge90,schacham92,leadley}
 The Fourier transform of the data versus inverse
magnetic
field yields two peaks, one at  a frequency $f_1=36.4\
\rm T$
corresponding to a density in the lowest subband
$n_{e_1}=2ef_1/h=1.76\times 10^{16}\ \rm m^{-2}$ ($h$ is
Planck's
constant) and one peak at the difference frequency $f_1-
f_2=33.6\ \rm T$.  This implies a density in the second
subband of $n_{e_2}=2ef_2/h=0.14\times 10^{16}\ \rm m^{-
2}$. The Fermi level at the onset of the
population of the second subband ($E_F=\pi\hbar^2
n_H/m^{*}=72~\rm meV$) is very close to the
distance between the bottom of the two subbands measured
by
photoluminescence: $E_2-E_1=75\ \rm meV$.\cite{eindhoven}
This is evidence that the observed beat in the
SdH oscillations is not the result of inhomogeneities in
the
sample. Our $\rho_{xx}$ data do not show any sign of a
superposition of SdH oscillations
with two frequencies as quite often observed in systems
with two occupied
subbands.\cite{cunningham88,coleridge90,leadley,vanhouten}
 The beat we see in the SdH oscillations is only observed
over a very
narrow range of densities ($n_H=(1.75\pm0.01)\times
10^{14}\ \rm m^{-2}$).
As the density is increased further by illumination, any
obvious signs of the second subband disappear and the SdH
oscillations are observed on a rising background which
becomes
stronger with increasing density (Fig. \ref{5tshots}(c)
and
(d)). At the same time, in the Hall effect, $\rho_{xy}$
is seen to
deviate from the classical straight line. As will be
discussed below, this
is evidence for a second type of carrier with a much
lower
mobility.

The density $n_{e_1}$ of the lowest subband in the
Ga$_{0.8}$In$_{0.2}$As quantum well, obtained from the
period of the SdH oscillations, is shown versus the Hall
density $n_H$ (see Eq. (\ref{eqnhall})) in Fig.
\ref{nsdh_tauq}(a). Below
$n_H=1.8\times 10^{16}\ \rm m^{-2}$ the
small difference between the data points and the dashed
$n_{e_1}=n_H$ line is the  result of experimental error,
rather than a consequence of the very small number of
carriers in the second subband around $n_H\approx
1.75\times 10^{16}\ \rm m^{-2}$. For densities $n_H \geq
1.85\times 10^{16}\ \rm m^{-2}$ however, a big difference
between the data points and the line is apparent.
This shows that carriers in the Ga$_{0.8}$In$_{0.2}$As QW
no longer account
for the total density beyond $n_H=1.85\times
10^{16}\ \rm m^{-2}$. At the same value for $n_H$ the
transport lifetime $\tau_t=\mu_H m^*/e$, shown in Fig.
\ref{nsdh_tauq}(b), starts to decrease  with increasing
Hall density.

\subsection{Multi-subband occupation and parallel
conduction}\label{sectwoband}

The resistivity of a multi-subband 2DEG or a 2DEG with a
parallel conduction layer in the presence of a magnetic
field
$B$ is usually described by the sheet resistivity tensor
$\tilde{\rho}_i$  where the subscript $i$ labels the
conducting layer.
If the conduction in layer $i$ with electron density
$n_i$ is
due to a process
characterized by a single relaxation time $\tau_{t,i}$
then
the Lorentz force leads to\cite{ashcroft}
\begin{equation}
\tilde{\rho}_i= \left( \begin{array}{cc} \rho_{o,i} &
B/n_i e
\\
-B/n_i e & \rho_{o,i} \end{array} \right) =
\left( \begin{array}{cc} \rho_{xx,i} & \rho_{xy,i} \\
-\rho_{xy,i} & \rho_{xx,i} \end{array} \right),
\end{equation}
where  $\rho_{o,i} = m^*_i/n_i e^2\tau_{t,i}$ and $m^*_i$
is
the effective mass of the carriers in layer $i$.
The total current in the case of a parallel connected
multilayer
system is the sum of the currents in the different
layers. Thus the total sheet conductivity is
the sum of the sheet conductivities of the separate
layers.\cite{kane} If the conductivity of  layer $i$ is
given
by
\begin{eqnarray}
\tilde{\sigma}_i & = & \frac{n_i e^2\tau_{t,i}}{m^*_i}
\frac{1}{1+\omega_{c}^{2}\tau_{t,i}^2}\left(
\begin{array}{cc}
1  & -\omega_c\tau_{t,i} \\
\omega_c\tau_{t,i} & 1 \end{array} \right)\\
 & = & \left( \begin{array} {cc} D_i & -A_i
\\ A_i & D_i \end{array} \right),
\end{eqnarray}
with $\omega_{c,i}=eB/m^*_i$, then the total conductivity
of a
system with two layers, labeled $a$ and $b$, is
\begin{equation}
\tilde{\sigma}_{tot}=\tilde{\sigma}_a+\tilde{\sigma}_b=
\left(
\begin{array}{cc}  (D_a +D_b) & - (A_a + A_b)  \\
(A_a + A_b) & (D_a+D_b)  \end{array} \right).
\end{equation}
After inverting the above equations we obtain for the
resistivity
\begin{equation}\label{pgx}
\rho_{xx} = \frac{D_a + D_b}{(D_a+D_b)^2 + (A_a+A_b)^2}
\end{equation}
and
\begin{equation}\label{pgy}
\rho_{xy} = - \frac{A_a + A_b}{(D_a+D_b)^2 +
(A_a+A_b)^2}.
\end{equation}
These equations can be simply written in terms of the
densities $n_a$, $n_b$ and the
mobilities
$\mu_a$, $\mu_b$ of the two layers.\cite{kane}
The expressions are simplified in the case of low
magnetic fields (such that $\mu_aB,\ \mu_bB \ll 1$) so
that \begin{eqnarray}
\rho_{xx}  =  \frac{1}{e}\frac{1}{n_a\mu_a+n_b\mu_b}
=  \frac{1}{e}\frac{1}{n_H\mu_H} \label{wiske}
\end{eqnarray}
and
\begin{eqnarray}
\rho_{xy}  =
\frac{B}{e}\frac{n_a\mu_a^2+n_b\mu_b^2}{(n_a\mu_a+n_b\mu_
b)^2}
=  \frac{B}{e}\frac{1}{n_H}.\label{asterix}
\end{eqnarray}
 Equations (\ref{wiske}) and (\ref{asterix}) relate the
Hall
density
$n_H$ and the Hall mobility $\mu_H$ to the densities
$n_a$
and
$n_b$ and the mobilities  $\mu_a$ and $\mu_b$.

 We have fitted Eqs. (\ref{pgx}) and (\ref{pgy}) to the
low
field measurements of $\rho_{xx}$ and $\rho_{xy}$ in
order to obtain
values for the electron density and mobility in the Si
$\delta$-layer ($n_{e_3}=n_b$ and $\mu_3=\mu_b$), and the
mobility $\mu_1=\mu_a$ in the first subband in the
GaAs/\-Ga$_{0.8}$In$_{0.2}$As/\-Ga$_{0.75}$Al$_{0.25}$As
QW.  Since the number of carriers in the
second subband $n_{e_2}$ is small and the transport
mobility  $\mu_2$ of these
carriers is usually a few times smaller than the mobility
of the electrons in
the first subband,\cite{vanhouten} we ignored the second
subband in our fit (in the resistivity calculations in
the
next section the second subband is included, but it is
found
that it has a negligibly small effect on the resistivity
values). We used the density in
the first subband $n_{e_1}=n_a$, determined from the
frequency of the SdH oscillations, as an input parameter.
A typical fit is shown in Fig. \ref{5tfit}. The agreement
between the classical picture and the background of both
$\rho_{xx}$
and $\rho_{xy}$ is very good. The resulting parameters
are shown in
Table \ref{jefke}. The values obtained for the density
$n_{e_3}$ and the mobility $\mu_3$ indicate that the Si
$\delta$-layer has a very high electron density and has a
mobility which is $\approx$50 times smaller than the
mobility in the first subband.
Both the high density and the low mobility, are typical
for
parallel conduction in an impurity layer such as the Si
$\delta$-layer in our case. From the values in Table
\ref{jefke}, it is
easily verified that the mobilities and densities obey
the
relation  $n_H\mu_H=n_{e_1}\mu_1+n_{e_3}\mu_3$ as implied
by Eq. (\ref{wiske}).

\subsection{Scattering times}

{}From the temperature dependence of the amplitude of the
SdH
oscillations (see Sec. \ref{sdhosc}) we determined the
effective mass to be $m^*=(0.058\pm 0.005)\times m_0$ at
the dark value of the density ($m_0$ is the rest mass of
the free electron).  This is lower than the experimental
values for the effective mass for GaAs/\-Ga$_{1-
x}$In$_{x}$As/\-Ga$_{1-y}$Al$_{y}$As systems found in the
literature, which show a wide spread: from $m^*=0.063
m_0$  (Ref. \onlinecite{luo88}) to $m^*=0.072 m_0$ (Ref.
\onlinecite{butov90}) for GaAs/\-
Ga$_{0.87}$In$_{0.13}$As/\-Ga$_{0.7}$Al$_{0.3}$As, and
from $m^*=0.067 m_0$ (Ref. \onlinecite{butov90}) to
$m^*=0.071 m_0$ (Ref. \onlinecite{brugger91}) for GaAs/\-
Ga$_{0.82}$In$_{0.18}$As/\-Ga$_{0.7}$Al$_{0.3}$As. The
difference might in part be due to the higher In content
in our structures.

When $m^*$ is known, the single particle scattering time
can
be determined from the decay of the SdH oscillations with
decreasing magnetic field by making a so called Dingle
plot of  $\ln(\Delta\rho\sinh X/2\rho_0 X)$ versus
inverse magnetic field, with $\Delta\rho$ the measured
amplitude of the oscillations. The slope of the resulting
plot  is then inversely proportional to the single
particle scattering time $\tau_q$.\cite{footnote}

The results of the Dingle analysis are shown in
Fig. \ref{nsdh_tauq}(b). We see that with increasing
carrier
density the single particle scattering time increases. By
contrast, the transport scattering time, also shown in
Fig. \ref{nsdh_tauq}(b), decreases with increasing
carrier density.\cite{footnote2} This apparent
contradictory behavior can be understood as follows.
 The single particle scattering rate counts every
scattering
event. However, the transport scattering time in high
density
GaAs/\-Ga$_{1-x}$In$_{x}$As/\-Ga$_{1-y}$Al$_{y}$As
structures is mainly determined by large
angle scattering events such as cluster scattering due to
the
non-uniform distribution of In in the Ga$_{1-
x}$In$_{x}$As\cite{luo} and
intersubband scattering.\cite{stormer}  The latter
becomes
more important when the density increases so that the
Fermi level lies in, or very close to the second subband.
Also cluster scattering increases with
increasing density.\cite{luo}  In contrast, small angle
scattering such as remote ionized
impurity scattering due to the ionized Si donors in the
$\delta$-layer is more effectively screened when the
carrier
density is large, and the remote ionized impurity
scattering
is also screened by additional carriers in the second
subband
and the $\delta$-layer.
The effect of a reduction in the small angle scattering,
which dominates the single particle scattering time,
outweighs the concurrent increase in the large angle
scattering rate. Consequently the
single particle scattering time in the first subband
increases
with density while the transport life time decreases.

We do not see any clear effect of the depopulation of the
second subband on the Hall mobility or the quantum life
time
in the first subband as has been observed in GaAs/Ga$_{1-
x}$Al$_{x}$As
heterojunctions with two occupied subbands.\cite{kusters}
This is likely to be due to the relatively low mobilities
in the  GaAs/\-Ga$_{0.8}$In$_{0.2}$As/\-
Ga$_{0.75}$Al$_{0.25}$As heterojunction compared to those
in GaAs/Ga$_{1-x}$Al$_{x}$As structures.

\section{LARGE MAGNETIC FIELDS}\label{highmf}

\subsection{Introduction}

At sufficiently high magnetic fields, $\rho_{xx}$ can
become
vanishingly small and $\rho_{xy}$ shows plateaux in
finite
ranges of the magnetic field when $E_F$ lies between two
separated Landau levels. This feature of electrical
transport
in high magnetic fields in 2D systems is called the
quantized
Hall effect \cite{vonklitzing,prange,girvin,qhebook} and
has been observed in a wide range of semiconductor
heterostructures.

Typical high field transport data shown in
Fig.~\ref{intr_large} for a Hall density of  $n_H = 1.9
\times 10^{16}\rm~m^{-2}$: above 37 T $\rho_{xx}$ becomes
very small for the lower temperature $T$=1.4~K, and there
is a corresponding quantized plateau in $\rho_{xy}$ at
12.5~k$\Omega$. The rising background in $\rho_{xx}$ at
fields below 30 T is due to parallel conduction in the Si
$\delta$-layer which freezes out at 1.4 K and fields
$B>30$~T.
For the higher temperature, $T=4.2$~K, there is still
some parallel conduction and the high field minimum in
$\rho_{xx}$ is at about 200 $\Omega/\Box$.

\subsection{Theory}

At high magnetic fields the conductivities of a 2D system
are given by\cite{afs}
\begin{mathletters}
\begin{eqnarray}
\sigma_{xx} &=& \frac{e^2}{\pi^2 \hbar} \sum_{N,s} \int
dE \left
(-\frac{\partial f(E)}{\partial E}\right )\nonumber\\
 & & \times \left
[\frac{\Gamma_{N}^{xx}}{\Gamma_{N,s}}\right ]^2
(\pi^2 l^2 \Gamma_{N,s} D_{\Gamma_{N,s}}(E))^2 ,
\label{sxx1}
\end{eqnarray}
and
\begin{equation}
\sigma_{xy} = - \frac{e}{B} \sum_{N,s}\int dE
f(E)D_{\Gamma_{N,s}}(E).
\label{sxy1}
\end{equation}
\end{mathletters}
Here $f(E)$ is the Fermi-Dirac distribution function,
$D_{\Gamma_{N,s}}(E)$ is the density of states for
electrons in Landau level $N$
with spin $s$, $\Gamma_{N,s}$ is the width of the Landau
level and $\Gamma_N^{xx}/\Gamma_{N,s}$ is a dimensionless
factor which depends on the type of scattering.  The
chemical potential $\mu$ is
determined through the condition of conservation of total
concentration
of electrons
\begin{equation}\label{density}
n_e = \sum_{N,s} \int dE f(E) D_{\Gamma_{N,s}}(E).
\end{equation}
 The sum in the above equations runs over the single
particle
states whose energies  are given by
\begin{equation}
E_{N,s} = (N+\frac{1}{2}) \hbar\omega_{c} + s\frac{1}{2}
g^{*}
\mu_B B ,
 \end{equation}
with $s=\pm 1$, $ g^{*}$ is the effective spin-splitting
factor and $\mu_B$ the Bohr magneton. The
spin-splitting factor for bulk GaAs is 0.44 (Ref.
\onlinecite{weisbuch77}) but it has been reported
that in a high magnetic field $g^{*}$  starts to
oscillate due to exchange enhancement, reaching values as
high as 2.5.\cite{englert2,nicholas}
For all the calculations we have used a constant value
for $g^*$=2.0 for the whole range of magnetic fields.
This value is larger than the $g$-factor at zero field
but it is near the measured values for this magnetic
field range.\cite{englert2}

The ratio
$\Gamma_{N}^{xx}/\Gamma_{N,s}$  depends on the type of
scattering.
For short-range scattering Ando {\it et al.}\cite{afs}
found:
\begin{equation}\label{gammanxx}
\left[\frac{\Gamma_{N}^{xx}}{\Gamma_{N,s}}\right]^2=
(N+\frac{1}{2}).
\end{equation}
Ando and Uemura \cite{andouemura} calculated this ratio
numerically for the case of a semi-elliptic DOS.
They found that for long-range scattering the ratio
 $(\Gamma_{N}^{xx}/\Gamma_{N,s})^2$
 is smaller than $(N+1/2)$ and the difference from the
short range scattering result increases with
the Landau level index $N$.

In order to compare our high magnetic field data with
theory
a Gaussian form for the total DOS  has been used
\begin{equation}\label{totdos}
D_{\Gamma_{N,s}}(E) = \frac{1}{2\pi
l^2}\frac{1}{\sqrt{2\pi}\Gamma_{N,s}}
 e^{-(E-E_{N,s})^2/2\Gamma_{N,s}^2},
\end{equation}
where  $l=\sqrt{\hbar /eB}$ is  the magnetic length and
$1/2\pi\l^2$ is the available number of states in each
Landau
level.
In the limit of short range scatterers Ando {\it et
al.}\cite{afs} found for the level broadening
$\Gamma_{N,s}$:
\begin{equation}
\Gamma_{N,s}^2 = \frac{2}{\pi} e^2
\frac{\hbar^2}{m^{*^{2}}}
\frac{B}{\mu_{q}},
\label{dsc}
\end{equation}
where $\mu_q=e\tau_q/m^*$ is the quantum mobility.

In the center of the Landau levels the electron states
are
extended while those in the tail are localized and
consequently do not contribute to the dissipative part of
the
conduction. In the phenomenological model of Englert
\cite{englert}  the tails of the DOS which do not
contribute to conduction are taken into account by using
a Gaussian DOS $D_{\lambda_{N,s}}(E)$ of {\em extended
states} with a width $\lambda_{N,s} < \Gamma_{N,s}$ (see
Fig.~\ref{dos}).
Substituting this DOS in Eq. (\ref{sxy1}) and Eq.
(\ref{sxx1})
results in
\begin{mathletters}
\begin{equation} \label{sxydos}
\sigma_{xy}\frac{h}{e^2} = -\sqrt{\frac{2}{\pi}}
\sum_{N,s}\int dE f(E)
\frac{1}{\lambda_{N,s}}
e^{-(E-E_{N,s})^2/2\lambda_{N,s}^2} ,
\end{equation}
and
\begin{eqnarray} \label{sxxdos}
\sigma_{xx}\frac{h}{e^2}& = &\frac{1}{4}
\sum_{N,s}
\Gamma_{N,xx}^2 (N+\frac{1}{2})\nonumber\\
 & & \times \int dE
\left(-\frac{\partial f(E)}{\partial E}\right)
e^{-(E-E_{N,s})^2/\lambda_{N,s}^2}.
\end{eqnarray}
\end{mathletters}

 In this last equation we introduced the factor
$\Gamma_{N,xx}^2=
(\Gamma_{N}^{xx} / \Gamma_{N,s})^2/(N+1/2)$, which equals
one
in the case
of short-range scatterers. In the evaluation of the Fermi
level all the electrons contribute and consequently
the total DOS
$D_{\Gamma_{N,s}}(E)$ is used.

The approximation of short range scatterers is clearly
not
satisfied in the modulation doped heterostructures of the
present study.
{}From  Ando {\it et al.} \cite{afs} it is obvious that for
long range
scattering the peak transverse conductivity decreases
rapidly
with increasing scattering range. Therefore in our
calculations we used  the ratio $\Gamma_{N,xx}$ as a
fitting
parameter in order to give information about the range of
the
scattering centers  relevant for the present samples.

With changing magnetic field   the population
of the different  Landau levels within one 2D layer
changes
and it is possible that the electrons   tunnel from the
QW to the $\delta$-layer
and vice versa.  This results in a Fermi energy which
depends on the magnetic field. At $T=0$
Eq.~(\ref{density}) becomes
\begin{equation}
n_e=\sum_{i} n_{e_i} = \sum_{i} \sum_{N,s} \int_{0}^{E_F}
dE D_{\Gamma_{N,s}}^{i}(E),
\end{equation}
where the index $i$ indicates the conducting layer.
 For low temperatures ($T < 10$~K)
we expect the Fermi energy to be very close to the  $T=0$
value.
For the $B=0$ case and for low temperatures we can
calculate the Fermi energy for a perfect 2DEG
from the number of  electrons using the relation
\cite{afs}
\begin{equation}
n_e= \frac{ m^{*}}{\pi \hbar^{2}} (E-E_F)
\label{eef}
\end{equation}
where $E$ is the bottom of the subband.

When the magnetic field becomes sufficiently large it
will
affect the equilibrium population of the free electrons
in the
$\delta$-layer. A magnetic field shrinks the electron
wavefunction, leading to an increase of the binding
energy of
the donor impurities and eventually into the freeze-out
of the
electrons in the impurity bound states \cite{yka,mares}.
For a
structure with a single type of carrier the onset of
the freeze-out regime corresponds to the magnetic field
at
which $\rho_{xy}$ starts to increase abruptly with
increasing $B$.
In our calculations we have obtained the threshold field
$B_{thr}$ from
the
experimental $\rho_{xy}$ data.
For magnetic fields beyond this threshold field we can
approximate the number of electrons by
\cite{beckman,amirkhanov}
\begin{equation}
n_e = n_0 e^{-\epsilon_B/k_B T}
\label{efr}
\end{equation}
where $n_0$ is the density below the freeze-out
threshold, and $\epsilon_B$ is the electron binding
energy which depends on  the  magnetic
field  through
\begin{equation}
\epsilon_B = b(B-B_{thr})^{1/3},
\end{equation}
and $b$ is a constant which is taken as a fitting
parameter.
This reduction of the number of
electrons results in a decrease of the Fermi energy and
consequently leads to a lowering of the
Landau level occupation and an increase in
$\rho_{xy}$.

\subsection{Comparison with experiment}
\subsubsection{The QHE in the Ga$_{0.8}$In$_{0.2}$As QW}

In Fig.~\ref{qhe50t} we show the measured resistivities
(full curves) $\rho_{xx}$ and $\rho_{xy}$  of the
unilluminated structure with an electron concentration of
$n_H=1.69\times 10^{16}~\rm m^{-2}$  at $T$=1.4 and 4.2
K.
At this density only the lowest subband in the GaAs/\-
Ga$_{0.8}$In$_{0.2}$As/\-Ga$_{0.75}$Al$_{0.25}$As quantum
well is occupied and $n_H$ equals the total electron
density
in the structure. There is no parallel conduction in the
$\delta$-layer and the system shows the QHE.  At 1.4 K
the
$\rho_{xy}$ plateau at a Landau level filling factor
$\nu=2$ extends
over a field range of 10 T and corresponds to a
$\rho_{xx} = 0$
minimum of the same width. Deep minima in $\rho_{xx}$ are
also seen at filling factors $\nu=4$ and $\nu=6$.

Figure~\ref{qhe50t} also shows the theoretical
calculations for high fields using the Englert model
described above (dotted curves).
The parameters used for the calculations are shown in
Table
\ref{table2}.
The agreement between the
theoretical calculations and the experimental data for
$\rho_{xx}$ and $\rho_{xy}$ is fairly good.
Notice (see Table II) that we took $\Gamma_{N,xx} = 0.65
< 1$, which indicates that there is a significant
contribution from small angle scattering. This is in
agreement with Fig.~\ref{nsdh_tauq} which shows that the
transport
relaxation time is almost an order of magnitude larger
than the quantum
lifetime, a characteristic feature of small angle
scattering.
This also agrees with the fact that the width of the
Landau levels $\Gamma_i$
is much smaller than the width $\Gamma_{\delta}$ we would
infer from a theory with delta-scatterers
($\Gamma_{i}/\Gamma_{\delta}= 0.3$, see Eq.~(\ref{dsc})).
For
$T$=1.4 K the agreement between theory and experiment is
rather good, for both $\rho_{xx}$ and $\rho{xy}$. For
$B>40\ \rm T$ the agreement is less satisfactory
indicating a smaller value for the width of the lowest
Landau level.
Note that in our analysis we assumed $\Gamma_i ,
\lambda_i$ and
$\Gamma_{N,xx}$ to be independent of the Landau level
index $N$ in order
to limit the number of fitting parameters.
It is known that for non-delta-scatterers this assumption
breaks down.\cite{afs}

A similar discrepancy between theory and experiment at
fields $>40~\rm T$ is seen for $T=4.2~\rm K$. The
observed spin-splitting in $\rho_{xx}$ at $B\approx 22
{}~\rm T$ harly shows up in the theoretical results,
indicating an exchange enhanced spin-splitting
factor\cite{englert2,nicholas} at this field which is
larger than the value $g^*=2.0$ we used in the
calculations.

\subsubsection{Parallel conduction}

In Fig.~\ref{1.84exp-theo} $\rho_{xx}$
and $\rho_{xy}$ are shown for a Hall density
$n_H=1.84\times
10^{16}~\rm m^{-2}$ at $T=1.4$ and 4.2 K. The parameters
used in the calculations are given in Table~\ref{table2}.
{}From Fig.~\ref{1.84exp-theo}  we notice that  there is a
reasonable agreement between theory and experiment for
1.4 K when we assume that the ground subband in the
$\delta$-layer is not populated.
However, at 4.2 K the experimental data clearly show
that: 1) the minima in $\rho_{xx}$ below 30~T are no
longer zero, and 2) $\rho_{xy}$ slightly deviates to
lower values than expected from an extrapolation
from the $B=0$ behaviour. Both experimental findings
indicate
that there is parallel conduction
in the $\delta$-layer, or in the second subband of the
QW.
This is in agreement with the results of
Sec.~\ref{sectwoband} and
Table {\ref{jefke}, where we found that for
 $n_H=1.87\times 10^{16}~\rm m^{-2}$
the ground subband in the $\delta$-layer is populated
with a density $n_{e_3} \approx 0.7 \times 10^{16}~\rm
m^{-2}$.  For $T=1.4$ K however, the magnetic freeze-out
of these carriers occurs already at small magnetic
fields. Therefore  those carriers are not included in the
calculation.
{}From Eq.~(\ref{efr})  it is obvious that at 4.2 K the
effect of the magnetic freeze-out is strongly weakened,
resulting in parallel conduction and $\rho_{xx}$ minima
which are no longer zero.
 But for $B>30$ T the electrons in the Si $\delta$-layer
are again practically frozen out so that
the $\rho_{xx}$ minima become almost zero and $\rho_{xy}$
exhibits a plateau at the quantized value of $h/2e^2$.

In Fig.~\ref{1.95exp-theo} (a) and (b) we show the
resistivities $\rho_{xx}$ and $\rho_{xy}$ at 1.4 K for
$n_H = 1.97\times10^{16}~\rm m^{-2}$.  The parameters
used
to obtain the theoretical curves (dashed/dotted lines)
are given in Table~\ref{1.95}. The experimental
$\rho_{xy}$ data  (solid line) in
Fig.~\ref{1.95exp-theo}(b) show a steep increase at $B=
36$ T
from 6.4~k$\Omega$ to 12.8~k$\Omega$, which coincides
with a
steep drop in $\rho_{xx}$
These  changes in the transport data can again be
attributed
to the magnetic freeze-out of the electrons in the
$\delta$-layer.
In the calculation we used $b=0.1~\rm meV/T^{1/3}$. We
used as threshold field $B_{thr}=36$ T. The binding
energy is of the order of 0.1 meV at $B=37$ T
at which point  it equals the
thermal energy $k_B T$ for $T=1.4$ K.
The electron single particle relaxation times differ by
almost two orders of magnitude for the ground subband of
the
QW as compared to the ground subband of the $\delta$-
layer.
Therefore  the field where the Isihara-Smr\u{c}ka
model breaks down is much higher for the $\delta$-layer
than for the QW. To take this into account in our
calculations we divided the high magnetic field range
($B>5$ T) into two different regimes: for magnetic fields
$B<27$ T the Englert model is used
for the ground subband of the QW where $\omega_c\tau_q
>1$, while the Isihara-Smr\u{c}ka  model is applied to
the ground subband of the $\delta$-layer.
At fields $B > 27$ T we applied  the Englert model for
both layers.
  The agreement between theory and experiment for both
$\rho_{xx}$ and
$\rho_{xy}$ is fairly good for fields lower than 25 T and
for fields
larger than 35 T.
 There are discrepancies in the area between 25 and 35 T,
which are probably due to the simple approach that we
used for the magnetic freeze-out model and the transition
between different theoretical models. We have also not
taken into account changes in the shape of the potential
well and in the transport and quantum mobilities which
are a consequence of the fast descent of the number of
the electrons in the $\delta$-layer due to the magnetic
freeze-out.

\subsubsection{The QHE in two layers}

Figures \ref{2.05exp-theo} (a) and (b) show $\rho_{xx}$
and $\rho_{xy}$ at 1.4 K for a Hall density of $n_H=2.05
\times 10^{16}~\rm m^{-2}$.
The corresponding Fermi level $E_F$ and the carrier
densities $n_{e_i}$ in
the different subbands $i$ are given in
Figs.~\ref{2.05exp-theo} (c) and (d). In this calculation
we used $m^*=0.067m_0$ and $0.058m_0$ for the electron
effective masses in GaAs and Ga$_{0.8}$In$_{0.2}$As
respectively. The $B=0$ values of the Fermi energy for
each subband are calculated from
Eq.~(\ref{eef}). The parameters we used in the
calculations and the Landau-level broadening in case of
$\delta$-scattering (Eq.~(\ref{dsc})) are shown in
Table~\ref{2.05}. In our calculations we have extended
the two-carrier model discussed in Sec.~\ref{sectwoband}
to the case of three conducting layers and we have used a
constant $\Gamma_{N,xx}$ = 0.75 for the QW for all the
Landau levels. For the $\delta$-layer we have used
$\Gamma_{N,xx}$ = 0.95
 for all Landau levels. This reflects the fact that in
the $\delta$-layer the scattering is predominantly due to
the large concentration of background impurities (Si-
donors) resulting in large angle scattering very similar
to $\delta$-scattering.

At $B=0$ there are two subbands occupied in the QW and
one in
the $\delta$-layer. As can be seen from Fig.
\ref{2.05exp-theo}(b) the second subband in the
Ga$_{0.8}$In$_{0.2}$As QW has a very low carrier
concentration compared to the other two layers; therefore
it contributes very little to the conductivities. We used
the following parameters for the second subband in our
calculations: $n_{e_2}= 0.05 \times 10^{16}~\rm m^{-2}$,
$\Gamma_2= 1.60 \sqrt{B}~\rm meV$, $\lambda_2= 1.25
\sqrt{B}~\rm meV$, $\mu_2= 0.3~\rm m^2/Vs$ and $\mu_{q_2}
= 0.1~\rm m^2/Vs$. These values are only accurate to
within a factor of three. This low accuracy in the
determination of these parameters is due to the fact that
because of the low electron density $n_{e_2}$ there is
practically no contribution of this layer to the
resistivities $\rho_{xx}$ and $\rho_{xy}$.

An oscillatory electron density in the different electron
layers is observed.
This is due to the pinning of the Fermi energy in the
$\delta$-layer combined with electron flow between the
two layers.
 At 37 T the lowest subband in the QW and the lowest
subband
in the $\delta$-layer have nearly equal densities. Due to
the
high electron density in the $\delta$-layer magnetic
freeze-out no longer occurs in the experimentally
accessible magnetic field range. However, the magnetic
field is strong enough to introduce Landau quantization
effects in the low mobility 2DEG in the $\delta$-layer.
As a result we have two parallel 2DEGs, each showing the
QHE
at a Landau level filling factor of 2 at $B\approx 37$ T,
which results in a total filling factor $\nu=4$.
Consequently a  deep minimum in $\rho_{xx}$ and a
quantized Hall
plateau at $\rho_{xy}=h/4e^2=6.5~\rm k\Omega$ are
observed.

As is apparent from Fig.~\ref{2.05exp-theo} a
satisfactory agreement between theory and experiment has
been obtained. The agreement is best at low magnetic
fields where the Isihara-Smr\u{c}ka model is used.
 For $B>35~\rm T$ there is a slight disagreement between
theory and experiment in the $\rho_{xy}$ results. This is
probably due to the fact that $\rho_{xy}$ is still
quantized despite $\rho_{xx} \neq 0$. This makes it
difficult to achieve a perfect agreement for both
$\rho_{xx}$ and $\rho_{xy}$ within our simple theoretical
framework.

\section{CONCLUSION}

The magnetoresistance $\rho_{xx}$ and the Hall resistance
$\rho_{xy}$ in a GaAs/\-Ga$_{0.8}$In$_{0.2}$As/\-
Ga$_{0.75}$Al$_{0.25}$As heterostructure with a Si
$\delta$-layer parallel to the Ga$_{0.8}$In$_{0.2}$As QW
were measured in magnetic fields up to 50~T. We provide a
coherent description of the transport phenomena using two
subbands in the Ga$_{0.8}$In$_{0.2}$As  QW and one
subband in the Si $\delta$-layer.

Comparing the low field data with the semi-classical two-
band
model we showed that in the illuminated structure the
$\delta$-layer has a high concentration of carriers with
a low
mobility. Changing the electron density through
illumination has a strong,
but opposite effect on the transport scattering time and
the
quantum scattering time.

For a theoretical description of $\rho_{xx}$ and
$\rho_{xy}$ for the complete experimental magnetic field
range, we used the Isihara-Smr\u{c}ka model to fit the
SdH oscillations in the low field regime where $\omega_c
\tau_q <1$. The {\em low field} range extends up to 30 T
for the Si $\delta$-layer while for the QW this range
does not extend beyond 5 T. For the high field regimes we
employed the
Englert model, which is basically a phenomenological
picture for the quantum Hall effect. For low carrier
densities in the $\delta$-layer we had to include
magnetic freeze-out to account for the loss of carriers
at high fields in this layer.
We combined these different models using the two band
model to
obtain values for $\rho_{xx}$ and $\rho_{xy}$. Using only
the level
broadening, the binding energy and the freeze-out
threshold field as fitting parameters we obtained
satisfactory agreement between the experimental data and
theory. For fields $B>37-40~\rm T$ the agreement is less
satisfactory which is probably due to the fact that we
assumd a level broadening independent of the Landau level
index.

Since the second subband in the Ga$_{0.8}$In$_{0.2}$As QW
has a very small
amount of carriers as compared to the first subband, this
contribution
to conduction is hard
to detect in transport experiments.   The remaining
discrepancies between theory and experiment can also be
partly
due to a lack of precise information about the
second subband and the subband structure in the Si
$\delta$-layer.

\section{AKNOWLEDGEMENTS}
M.v.d.B and V.C.K. acknowledge the support of the Human
Capital and Mobility programme of the EC. F.M.P. is
supported by the Belgian National Fund for Scientific
Research. This work was also supported by a cooperation
programme of the Inter-university Microelectronics Center
(IMEC) and the Flemish Universities.

\begin{figure}
\caption{The structure of the studied pseudomorphic
GaAs/\-Ga$_{0.8}$In$_{0.2}$As/\-Ga$_{0.75}$Al$_{0.25}$As
system.
At 5 nm from the Ga$_{0.8}$In$_{0.2}$As QW a Si $\delta$-
layer is grown.}
\label{sample}
\end{figure}

\begin{figure}
\caption{The structure of the conduction band ($E_C$)
under
different illumination conditions. In the unilluminated
sample
(a), just one subband $E_1$ in the Ga$_{0.8}$In$_{0.2}$As
well is occupied.
When the structure is illuminated, the second subband
$E_2$ in
the Ga$_{0.8}$In$_{0.2}$As well becomes populated (b).
Further illumination
brings the lowest level $E_\delta$ in the $\delta$-layer
below
the Fermi level $E_F$ and $E_2$ depopulates (c). When the
density is satured, both subbands in the
Ga$_{0.8}$In$_{0.2}$As QW and
$E_\delta$ in the $\delta$-layer are occupied (d).}
\label{ec}
\end{figure}

\begin{figure}
\caption{Low field data for $\rho_{xx}$  (full line) and
$\rho_{xy}$
(dotted line) for four different Hall densities:
$n_H=1.67$
(a), 1.75 (b), 1.97 (c) and $2.21\times 10^{16}~\rm m^{-
2}$
(d). The dashed line represents the classical Hall
resistance for a single carrier system with density
$n_H$.}
\label{5tshots}
\end{figure}

\begin{figure}
\caption{(a) The electron density $n_{e_1}$ in the lowest
subband in the GaAs/\-Ga$_{0.8}$In$_{0.2}$As/\-
Ga$_{0.75}$Al$_{0.25}$As QW ($\Box$) versus the
Hall density $n_H$. The dashed line represents the
$n_{e_1}=n_H$ line. (b) The transport lifetime $\tau_t$
($\bullet$) and the quantum life time $\tau_q$ in the
first subband in the
Ga$_{0.8}$In$_{0.2}$As QW ($\diamond$) versus the Hall
density $n_H$.}
\label{nsdh_tauq}
\end{figure}

\begin{figure}
\caption{Fit of the two carrier model to the low field
$\rho_{xx}$
and $\rho_{xy}$ data at 4.2 K. The Hall density is
$n_H=2.07\times
10^{16}~\rm m^{-2}$. The dashed line represents the fit
with
parameters: $n_{e_1}=1.95\times 10^{16}~\rm m^{-2}$,
$\mu_1=2.8~\rm m^2/Vs$, $n_{e_3}=1.5\times 10^{16}~\rm
m^{-2}$ and
$\mu_3 = 0.06~\rm m^2/Vs$}
\label{5tfit}
\end{figure}

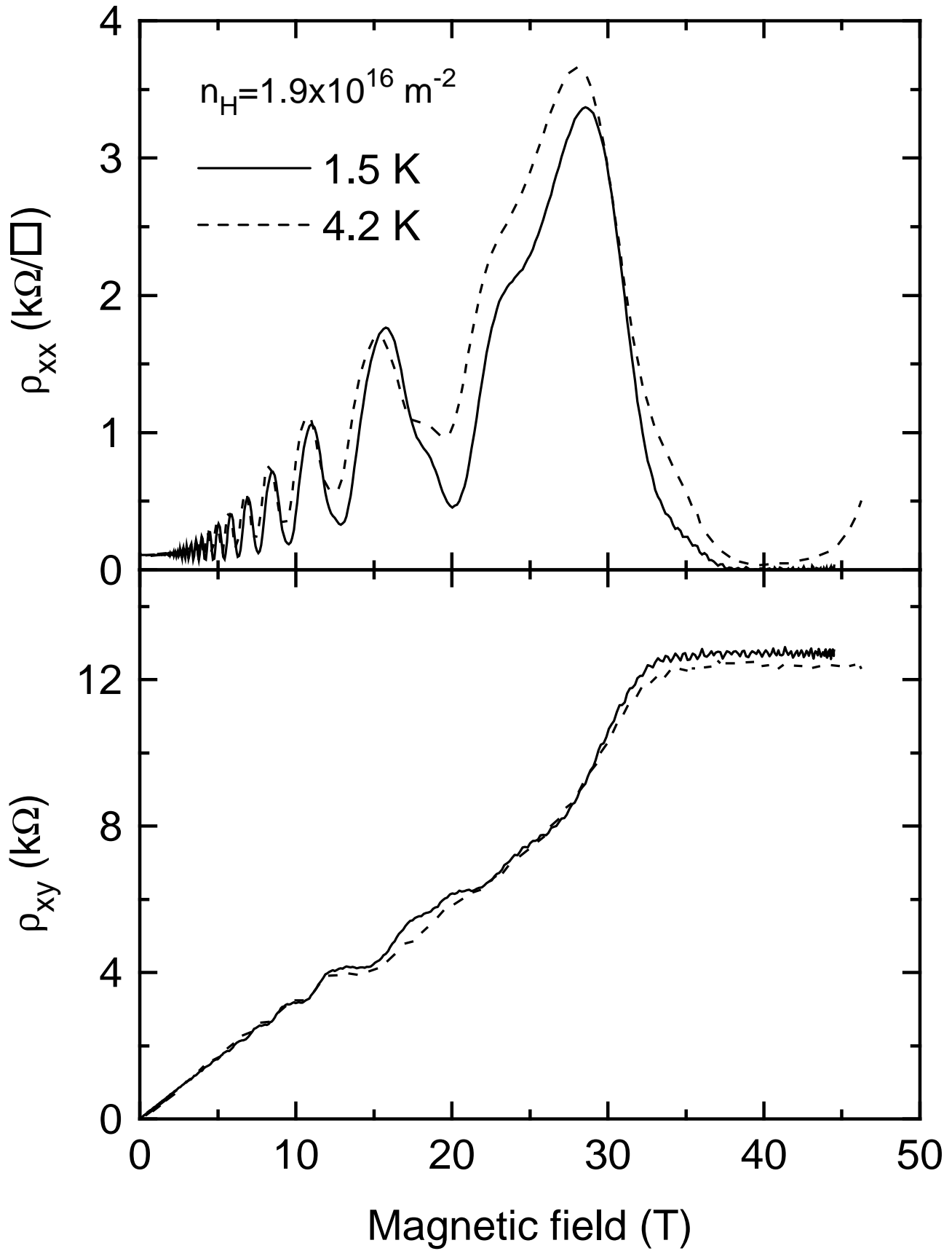
\begin{figure}
\caption{The high field data for $\rho_{xx}$ and
$\rho_{xy}$ at 1.4 K (full line)  and at 4.2 K  (dashed
line)
at a Hall density of $n_{H}=1.9\times
10^{16}~\rm m^{-2}$.}
\label{intr_large}
\end{figure}

\begin{figure}
\caption{The gaussian {\em total} density of states
$D_{\Gamma_{N,s}}(E)$ with width $\Gamma_{N,s}$ and the
density of {\em extended} states $D_{\lambda_{N,s}}(E)$
with width $\lambda_{N,s}<\Gamma_{N,s}$ used to calculate
the resitivities in the quantized Hall regime.}
\label{dos}
\end{figure}

\begin{figure}
\caption{$\rho_{xx}$ and $\rho_{xy}$ at 1.4 K ((a) and
(b)) and at 4.2 K
((c) and (d)) at a Hall density of $n_{H}=1.69\times
10^{16}~\rm m^{-2}$. The  full lines represent the
experimental data and the dotted line shows the
theoretical results.}
\label{qhe50t}
\end{figure}

\begin{figure}
\caption{$\rho_{xx}$ and $\rho_{xy}$ at 1.4 K ((a) and
(b)) and at 4.2 K
((c) and (d)) at a Hall density of $n_H=1.84\times
10^{16}~\rm
m^{-2}$. The  full lines represent the experimental data
and
the dotted line shows the theoretical calculations. No
magnetic freeze-out is included in these calculations.}
\label{1.84exp-theo}
\end{figure}

\begin{figure}
\caption{$\rho_{xx}$ (a) and  $\rho_{xy}$ (b) at 1.4 K at
a Hall density
of $n_H=1.95\times 10^{16}~\rm m^{-2}$; comparison
between
experiment (full line) and theory (dashed line in the low
field region where the Isihara-Smr\u{c}ka model is used
for
both the 2DEGs;
dotted line for the regime where the 2DEG in
the QW is described by the Englert model and  the 2DEG in
the
$\delta$-layer is described by the Isihara-Smr\u{c}ka
model;
dotted-dashed line for the regime where both 2DEGs are
described by the Englert model).
}
\label{1.95exp-theo}
\end{figure}

\begin{figure}
\caption{$\rho_{xx}$ (a) and  $\rho_{xy}$ (b) at 1.4 K at
a Hall density
of $n_H=2.05\times 10^{16}~\rm m^{-2}$; comparison
between
experiment (full line) and theory (dashed line in the low
field region where the Isihara-Smr\u{c}ka model is used
for
both the 2DEGs;
dotted line for the regime where the 2DEG in
the QW is described by the Englert model and  the 2DEG in
the
$\delta$-layer is described by the Isihara-Smr\u{c}ka
model;
dotted-dashed line for the regime where both 2DEGs are
described by the Englert model).
(c) shows  the  Landau levels and the Fermi energy $E_F$
and
(d) shows the electron densities in the different layers
(full line and dotted line for first and second subband
in the
QW; dashed line for the 2DEG in the $\delta$-layer.}
\label{2.05exp-theo}
\end{figure}

\begin{table}
\squeezetable
\caption{\label{jefke}Densities and mobilities obtained
from
the low  field transport data ($n_H$, $\mu_H$, $n_{e_1}$)
and
from the fit to the two-band model ($\mu_1$, $n_{e_3}$
and $\mu_3$).}

\begin{tabular}{c c c c c c}
$n_H$&$\mu_H$ & $n_{SdH}=n_1$ & $\mu_1$ & $n_2$ &
$\mu_2$\\
($10^{16}~\rm m^{-2}$) &($\rm  m^2/Vs$) &($10^{16}~\rm
m^{-
2}$) & ($\rm  m^2/Vs$) &($10^{16}~\rm m^{-2}$) &($\rm
m^2/Vs$)  \\  \hline
1.67  & 3.24  & 1.66  & 3.31  & -   & -  \\
1.87  & 3.25  & 1.84  & 3.32  & 0.7 & 0.02 \\
1.97  & 3.02  & 1.88  & 3.22  & 0.8 & 0.08  \\
2.07  & 2.65  & 1.95  & 2.81  & 1.5 & 0.06 \\
2.21  & 2.50  & 2.01  & 2.74  & 2.4 & 0.06 \\
2.53  & 1.98  & 2.22  & 2.17  & 3.9 & 0.07 \\
\end{tabular}
\end{table}

\begin{table}
\caption{\label{table2} Parameters for the two lowest
electron concentrations ($n_H=1.69\times 10^{16}~\rm m^{-
2}$,  Fig.~\protect{\ref{qhe50t}} and $n_H=1.84\times
10^{16}~\rm m^{-2}$, Fig.~\protect{\ref{1.84exp-theo}})
when only the lowest subband in the
Ga$_{0.8}$In$_{0.2}$As QW (index=1) is occupied.}
\begin{tabular}{l c c}
 & Fig.~\ref{qhe50t}& Fig.~\ref{1.84exp-theo} \\
 index   & 1  &  1    \\ \hline
\multicolumn{3}{c} {fitting parameters} \\ \hline
$\Gamma_i$ (meV) &  0.8$\sqrt{B}$  & 0.8$\sqrt{B}$ \\
 $\lambda_i$ (meV) &  0.2$\sqrt{B}$  & 0.24$\sqrt{B}$  \\
$\Gamma_{N,xx}$  & 0.65  &  0.8  \\
\hline
\multicolumn{3}{c} {physical parameters} \\ \hline
$\Gamma_\delta$ (meV) &  2.46$\sqrt{B}$  & 2.46$\sqrt{B}$
\\
$\mu_i~\rm (m^2/Vs)$    &  3.31      &  3.32  \\
$\mu_{q_i}~\rm  (m^2/Vs)$  & 0.31      & 0.31 \\
$n_{e_i}~\rm (\times 10^{16}~m^{-2})$   & 1.63  & 1.8 \\
\end{tabular}
\end{table}

\noindent\begin{table}
\caption{\label{1.95} Fitting parameters for the
theoretical
calculations in Fig.~\protect{\ref{1.95exp-theo}}
($n_H=1.95\times 10^{16}~\rm m^{-2}$). The indices
1 and 3 refer to  the first subband in the
Ga$_{0.8}$In$_{0.2}$As QW and the lowest
subband in the Si $\delta$-layer respectively.}
\begin{center}
\begin{tabular}{  l  c  c   }
index & 1 &  3 \\ \hline
\multicolumn{3}{c} {fitting parameters} \\ \hline
$\Gamma_i$ (meV) &   1.2$\sqrt{B}$&  3.50$\sqrt{B}$ \\
$\lambda_i$ (meV) &   0.8$\sqrt{B}$  &   2.3$\sqrt{B}$ \\
$\Gamma_{N,xx}$  & 0.75  & 0.9  \\
\hline
 \multicolumn{3}{c} {physical parameters} \\ \hline
$\Gamma_\delta$ (meV) &  2.27$\sqrt{B}$  & 7.3$\sqrt{B}$
\\
$\mu_i~\rm (m^2/Vs)$      &  2.82 & 0.05     \\
$\mu_{q_i}~\rm (m^2/Vs)$    & 0.38 &  0.048     \\
$n_{e_i}~\rm  (\times 10^{16}~m^{-2})$ & 1.89 & 0.9 \\
\end{tabular}
\end{center}
\end{table}

\noindent\begin{table}
\caption{\label{2.05} Fitting and physical parameters for
the theoretical calculations in
Fig.~\protect{\ref{2.05exp-theo}} ($n_H=2.05\times
10^{16}~\rm m^{-2}$). The indices
1 and 3 refer to  the first subband in the
Ga$_{0.8}$In$_{0.2}$As QW and the lowest
subband in the Si $\delta$-layer respectively.}
\begin{center}
\begin{tabular}{  l  c  c    }
index &  1 &   3  \\ \hline
 \multicolumn{3}{c} {fitting parameters} \\ \hline
$\Gamma_i$ (meV) & 1.4 $\sqrt{B}$ & 3.5$\sqrt{B}$ \\
$\lambda_i$ (meV) &   0.95
$\sqrt{B}$  &    2.25$\sqrt{B}$ \\
$\Gamma_{N,xx}$  & 0. 75 &  0.95  \\
\hline
 \multicolumn{3}{c} {physical parameters} \\ \hline
$\Gamma_\delta$ (meV) &  2.2$\sqrt{B}$  &  7.90$\sqrt{B}$
\\
$\mu_i ~\rm (m^2/Vs)$      &  3.00 &  0.05     \\
$\mu_{q_i} ~\rm (m^2/Vs)$    & 0.4 & 0.04     \\
$n_{e_i} ~\rm (\times 10^{16}~m^{-2})$   & 1.85 & 1.65 \\
\end{tabular}
\end{center}
\end{table}


\begin{references}
\bibitem{footnote}
Strictly speaking, this treatment is only valid in the
case
that one subband is occupied. However, we can also carry
out this Dingle analysis  for larger densities such
that the second subband and/or the $\delta$-layer are
occupied,
and when no intermodulation of the subbands is observed.
This is justified under the assumption that the decay of
the SdH oscillations from
the first subband is only determined by the single
particle
scattering time in this subband, and is independent of
the
featureless background
in
$\rho_{xx}$ due to the other conduction channels.
\end{references}
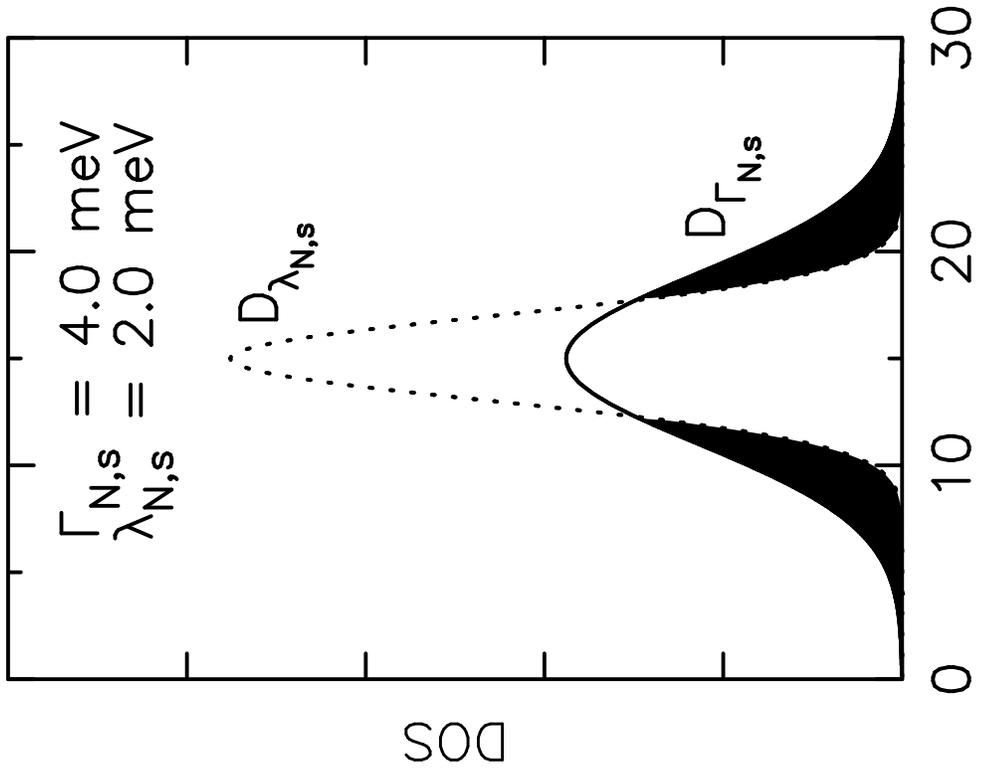
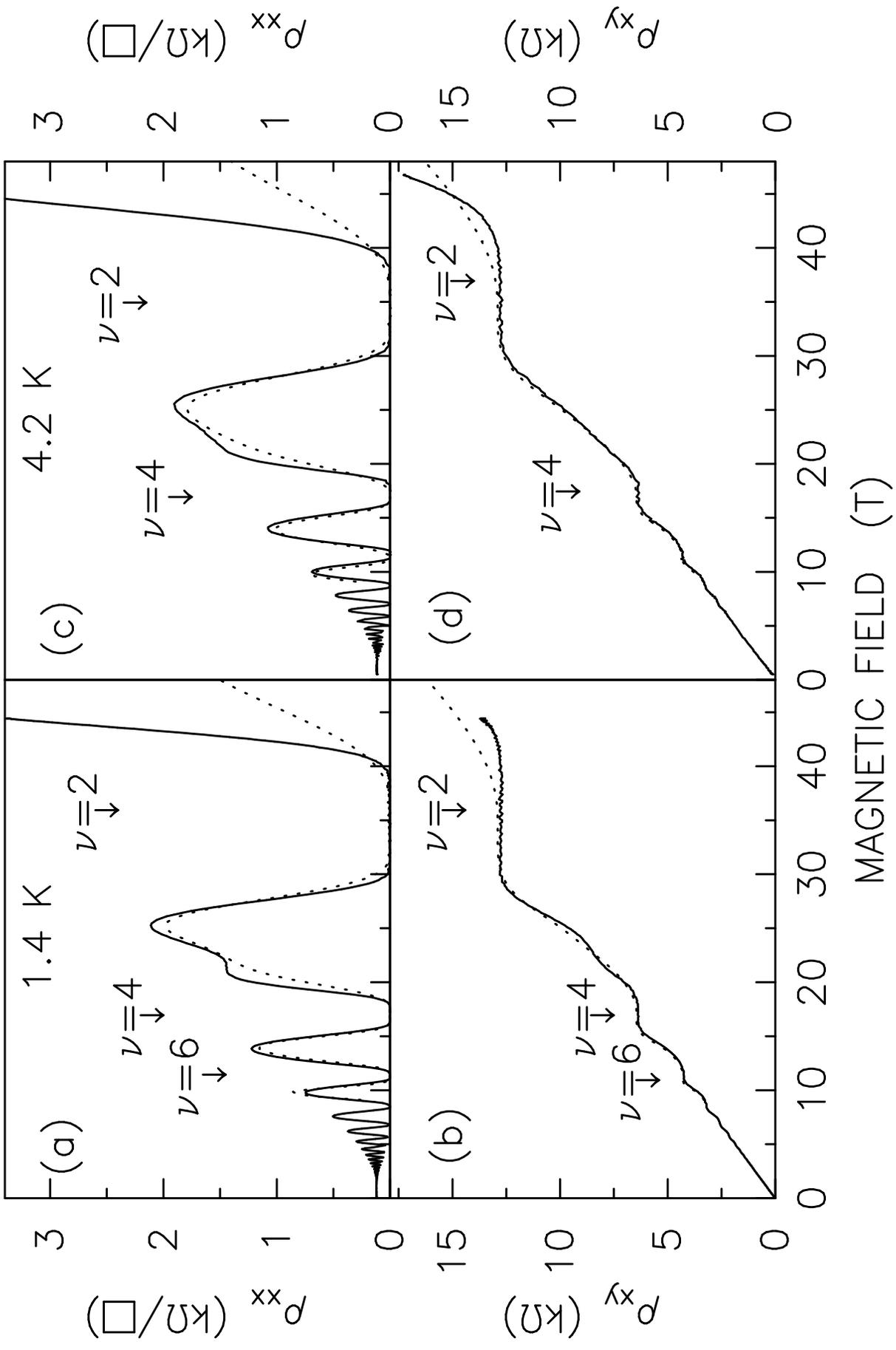
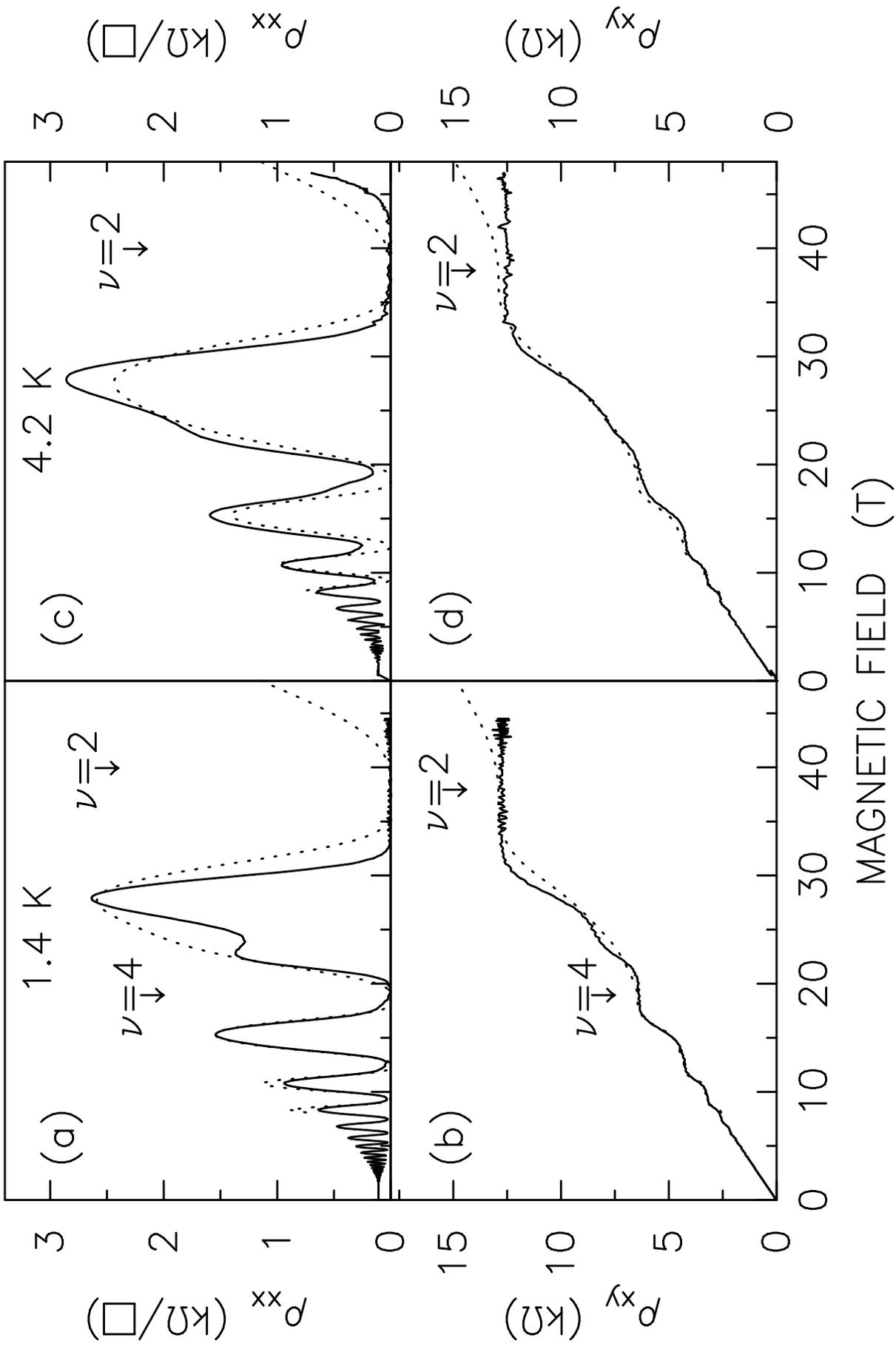
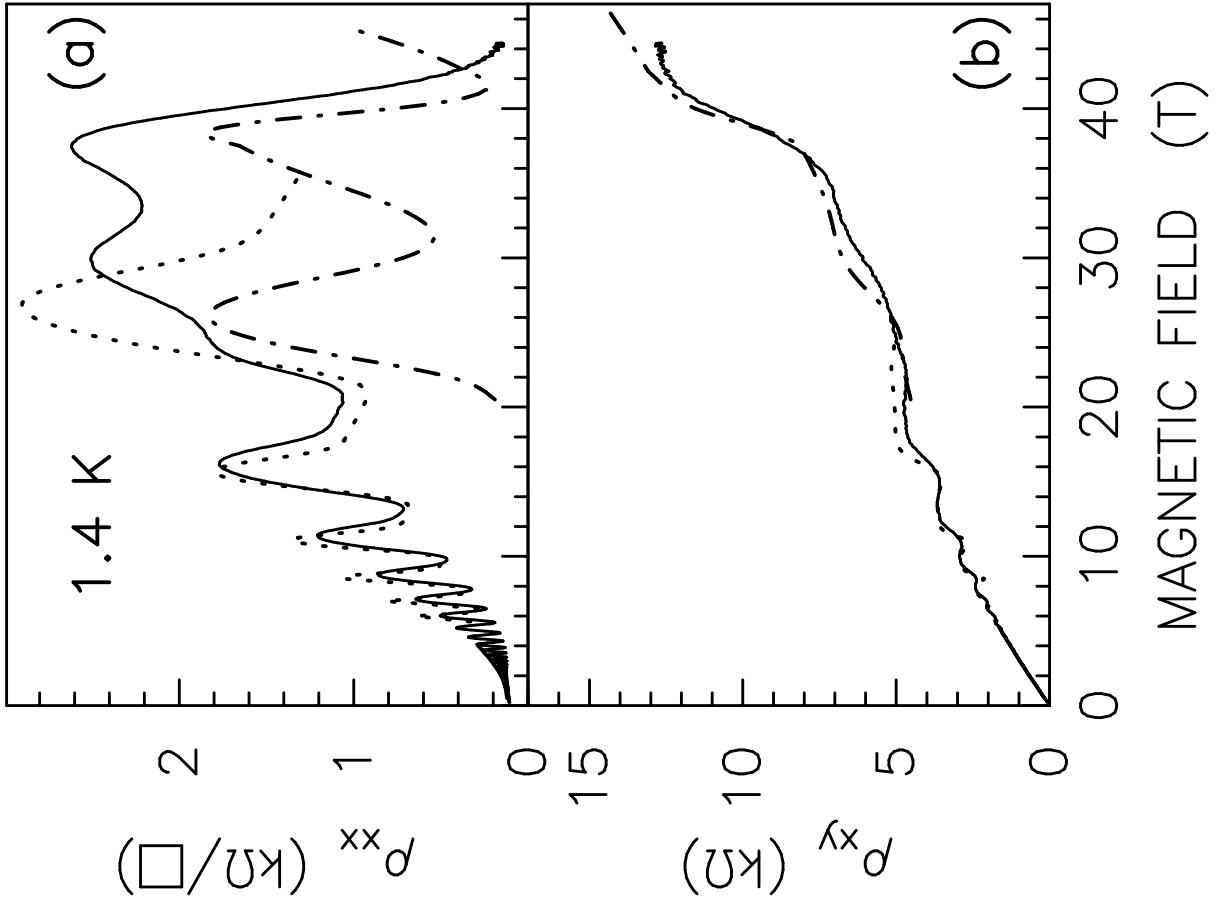
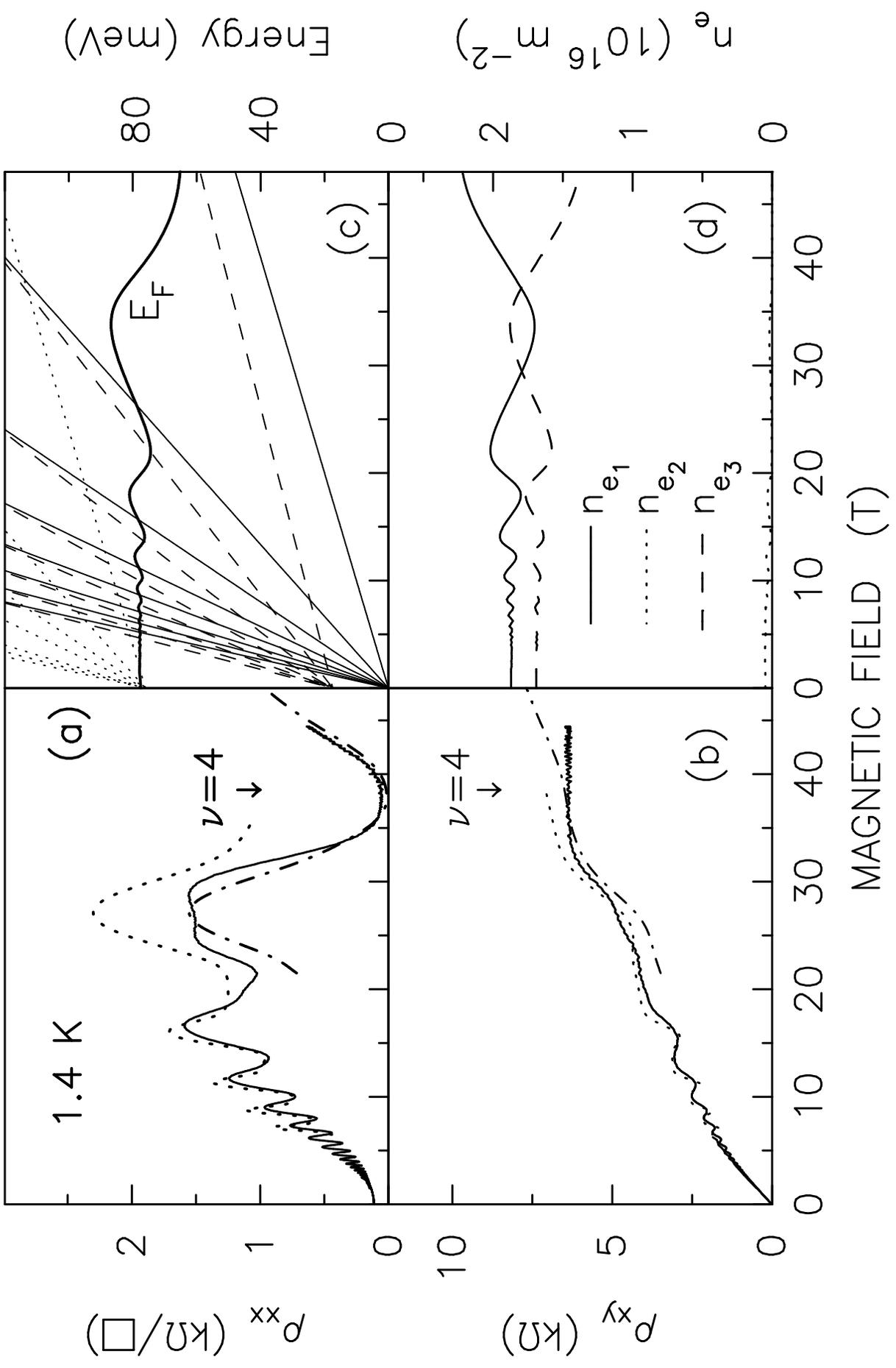
\end{document}